\documentclass[%
reprint,
amsmath,
amssymb,
aps,
pra,
superscriptaddress,
preprintnumbers,
prstab,
prstper,groupedaddress,,superscriptaddress]{revtex4-2}
\usepackage{graphicx}
\usepackage{dcolumn}
\usepackage{bm}

\usepackage[english]{babel}
\usepackage[utf8]{inputenc}
\usepackage{amssymb,amsfonts,amsmath,mathtools,mathrsfs}
\usepackage{relsize}
\usepackage{mdframed}
\usepackage{enumitem}
\usepackage{graphicx}

\usepackage[table,usenames,dvipsnames]{xcolor}
\usepackage{gensymb}
\usepackage{grffile}
\usepackage{float}

\usepackage[colorlinks=true,linkcolor=blue,citecolor=blue,urlcolor=blue]{hyperref}

\usepackage{textcomp}
\usepackage{braket} 

\begin{document}

\title{Vacancy induced expansion of spin-liquid  regime in $J_1-J_2$ Heisenberg model}

\author{Soumyaranjan Dash}
\affiliation{Department of Physical Sciences, Indian Institute of Science Education and Research (IISER) Mohali, Sector 81, S.A.S. Nagar, Manauli PO 140306, India}

\author{Anish Koley}
\affiliation{Department of Physical Sciences, Indian Institute of Science Education and Research (IISER) Mohali, Sector 81, S.A.S. Nagar, Manauli PO 140306, India}
\affiliation{Institut f\"ur Theoretische Physik, Universit\"at Leipzig, 04103 Leipzig, Germany}

\author{Sanjeev Kumar}
\email{sanjeev@iisermohali.ac.in}
\affiliation{Department of Physical Sciences, Indian Institute of Science Education and Research (IISER) Mohali, Sector 81, S.A.S. Nagar, Manauli PO 140306, India}

\begin{abstract}

We study the model for spin-1/2 $J_1-J_2$ Heisenberg antiferromagnets on a square lattice in the presence of spin vacancies. In order to overcome the methodological challenges associated with analyzing models with magnetic frustration and inhomogeneities, we introduce a new semi-classical approach in which singlet dimers are treated as effective classical degrees of freedom. The energetic and entropic aspects of the dimer formation are included via a classical Monte Carlo scheme that allows for the dynamical conversion of spin pairs into dimers and vice versa. We show that our semi-classical approach recovers the qualitative physics of the $J_1-J_2$ model in the absence of vacancies. The vacancies lead to a broadening of the spin-liquid regime between the N\'eel and the stripe antiferromagnetic phases. This suggests a possible new route to discover spin-liquid ground states by tuning the $J_2/J_1$ ratio in doped square lattice antiferromagnets. 
\end{abstract}

\date{\today}
\maketitle

{\it Introduction}:
The search for quantum spin-liquid phases in materials and models has been one of the heavily investigated topics in physics in recent years \cite{Cava2020, Zhou2017, Savary2016, Balents2010}. The possibility of discovering novel quantum correlated states with applications in quantum computing has rendered this research direction as one of the most relevant in the present times. Quantum spin models with competing interactions, which describe a broad class of magnetic materials, are commonly investigated for the possibility of spin-liquid (SL) ground states. One of the standard models in this context is the antiferromagnetic spin-$1/2$ $J_1$-$J_2$ Heisenberg model on square lattice. The existence and nature of spin-liquid phase near $J_2/J_1 = 0.5$ in this model continues to be a topic of active debate \cite{Yu2012b, Jiang2012, Morita2015, Richter2009, Richter2015, Qian2024}. Early studies suggested that a spin-liquid ground state exists in the range $0.4<J_2/J_1<0.6$ \cite{Liu2018, Hu2013, Li2012}. The possibility of a spin-liquid as well as a valence bond solid state (VBS) in the intermediate $J_2/J_1$ regime has also been reported \cite{Morita2015, Wang2018, Ferrari2020}. In the existing literature, the range of existence of the spin-liquid state appears to depend on the choice of method \cite{Hu2013, Morita2015, Gong2014,Mezzacapo2012, Sushkov2001}(see Supplemental Material). In fact, a recent study claims that there is no spin-liquid ground state in this model \cite{Qian2024}. 

The $J_1-J_2$ model in the small, intermediate, and large $J_2/J_1$ limits has proven useful in describing the experimental data on various magnetic compounds \cite{Coldea2001, Carretta2002, Melzi2000, Bombardi2004}. It has also been shown that the ratio $J_2/J_1$ can be tuned experimentally with the help of pressure \cite{Pavarini2008}. The presence of inhomogeneities, although ubiquitous in real materials, is typically viewed as a nuisance for both experiments and theory.  However, there are proposals that randomness can, in fact, be useful in inducing a spin-liquid ground state \cite{Uematsu2018, Savary2017, Yamaguchi2017, Potter2012}. On the theoretical side, the study of models with impurities or vacancies is known to be highly challenging, since the methods that utilize translational invariance cannot be employed for such problems. This leaves us with methods that explicitly treat spatial inhomogeneities, such as Monte Carlo (MC). However, quantum MC is known to suffer from the notorious sign problem for frustrated interactions such as those present in the antiferromagnetic $J_1-J_2$ model. The inaccessibility of suitable methods for studying the important problem of vacancies in a frustrated spin model serves as a motivation for us to implement a new approach. 

In this work, we introduce a semiclassical approach to develop a qualitative understanding of the ordered and disordered phases of quantum antiferromagnets at low temperatures ($T$). Our approach has a wide scope of applicability, and it is particularly useful for models with short-range interactions and disorder. We test our approach for the $J_1-J_2$ model on a square lattice without vacancies and show that the results are qualitatively similar to those obtained via other methods. The presence of vacancies leads to a broadening of the SL regime. Within our approach, we also identify vacancy-ordered states at low temperatures. Our results suggest a possible new route to induce SL phases in magnets that are realizations of the square-lattice $J_1-J_2$ model. We discuss the conditions and candidate materials for the potential realization of a doping-induced SL ground state.

{\it Model Hamiltonian and the Monte Carlo approach:}
We introduce the method by explicitly demonstrating its implementation on the spin-1/2 $  J_1$-$J_2 $ Heisenberg model on a square lattice, specified by the Hamiltonian,
\begin{eqnarray}
	H & = & J_{\text{1}} \sum_{\langle ij \rangle} {\bf s}_i \cdot {\bf s}_j + J_{\text{2}} \sum_{\langle \langle ij \rangle \rangle} {\bf s}_i \cdot {\bf s}_j , 
	\label{eq:Ham1}
\end{eqnarray}
\noindent
where $J_1$ and $J_2$ are the antiferromagnetic coupling constants between quantum spin-$1/2$ variables ${\bf s}_i, {\bf s}_j$ on the nearest neighbor (nn) and the next nn (nnn) sites, respectively. It is well known that the small and large $J_2/J_1$ limits of the model lead to ground states with conventional magnetic ordering that can be qualitatively captured within a classical approximation for the spins. Indeed, a classical Monte Carlo simulation of the model results in a $(\pi,\pi)$ N\'eel antiferromagnetic ground state for $J_2/J_1 < 0.5$ and a $(\pi,0)$ or $(0,\pi)$  stripe antiferromagnetic ground state for $J_2/J_1 > 0.5$.  The inclusion of quantum effects away from $J_2/J_1 = 0.5$, leads to a reduction in the ordered moment. The most significant change to this classical picture, upon including quantum correlations, occurs near $J_2/J_1 = 0.5$ where the energies of the two classical phases are in close vicinity. Most of the calculations support the existence of a regime with no conventional magnetic order in the parameter window $0.4 < J_2/J_1 < 0.6$  \cite{Hu2013, Liu2018, Li2012, Wang2018, Mezzacapo2012, Sushkov2001, Gelfand1990, Ferrari2020, Gong2014, Capriotti2000, Darradi2008, Sirker2006}. One of the simplest approaches that captures this quantum character of the ground state near $J_2/J_1 = 0.5$ is the cluster mean-field theory (CMFT) \cite{Ren2014}. However, the cluster approach leads to a valence-bond solid state due to the presence of cluster periodicity by construction.

By combining the features of a CMFT solution with the classical treatment for spins, we introduce a semiclassical Hamiltonian, $H_{\rm sc}$, given by,
\begin{eqnarray}
	H_{\rm{sc}} & = & J_{\text{1}} \sum_{\langle ij \rangle \in U} {S}_i {S}_j + J_{\text{2}} \sum_{\langle \langle ij \rangle \rangle \in U} {S}_i {S}_j \nonumber \\
 & & + E_b(J_1) \sum_{\langle  ij \rangle \in C} b_{ij} + E_b(J_2) \sum_{\langle \langle ij \rangle \rangle \in C} b_{ij}. 
	\label{eq:Ham2}
\end{eqnarray}


\noindent
The sites on the square lattice are dynamically separated into two categories: uncorrelated (U) and correlated (C). Spins on U sites are treated as classical variables, denoted as ${S}_i$. Correlated sites do not host local magnetic moments. However, two correlated sites linked via $J_1$ or $J_2$ represent the presence of a singlet between the corresponding pairs of sites. Therefore, we introduce $b_{ij}$ as binary bond variables that take values $0$ ($1$) corresponding to the absence (presence) of a singlet between spins at sites $i$ and $j$. The energy corresponding to bond variables at finite $T$ is given by,

\begin{eqnarray}
E_b(J_{\rm n}) = \frac{-3J_{\rm n}}{4}\left( \dfrac{1-e^{-J_{\rm n}/T}}{1+3e^{-J_{\rm n}/T}} \right),
\label{eq:Eb}
\end{eqnarray} 
\noindent 
where $k_{\rm B}$ is set to 1 and $n=1, 2$. The classical dynamics of the resulting Hamiltonian Eq. (\ref{eq:Ham2}) can be efficiently simulated within a Monte-Carlo approach. There are further constraints that we explicitly implement in the dynamics(see Supplemental Material). A similar semiclassical approach has recently been introduced by two of the authors in the context of Kondo lattice model \cite{Dash2025}. We compute the spin structure factor (SSF), the specific heat($C_\textrm{V}$), and the fraction of dimers($n_\textrm{D}$) in order to identify the phases and the boundaries between phases (see Supplemental Material).

 \begin{figure}
\includegraphics[width=0.96 \columnwidth,angle=0,clip=true]{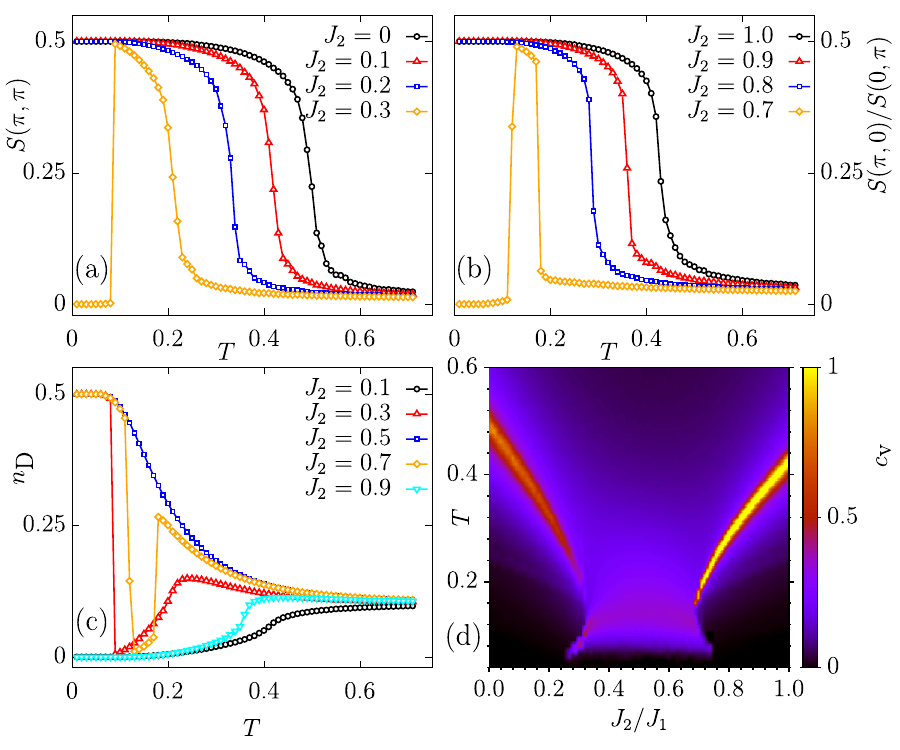}
 \caption{(Color online) Temperature dependence of, (a) $S(\pi,\pi)$, (b) $Max\{ S(\pi,0), S(0,\pi)\}$, and (c) the fraction of dimers, $n_\textrm{D}$, for different values of $J_2/J_1$. (d) The colormap of specific heat in $T- J_2/J_1$ plane.}
\label{fig1}   
\end{figure}

{\it Phase diagram for the $J_1-J_2$ model:--} 
To begin with, we test the performance of our new approach for the conventional $J_1-J_2$ model without vacancies. For small values of $J_2/J_1$, a N\'eel antiferromagnetic ground state is obtained, as inferred from the $T$ dependence of $S(\pi,\pi)$ (see Fig. \ref{fig1}(a)). The inflection point in the $T$ dependence of $S(\pi,\pi)$ provides a good estimate for the ordering temperature. For $J_2/J_1=1$, we obtain the stripe AFM ground state as inferred from the $(\pi,0)$ and $(0,\pi)$ components of the SSF (see Fig. \ref{fig1}(b)). The magnetic order vanishes as one approaches the classical transition point, $J_2/J_1 = 0.5$, from either limit. The presence of a non-magnetic ground state in the intermediate parameter regime is an important feature of the quantum problem that we effectively capture in our semi-classical approach. The ground state in the intermediate regime is characterized by the presence of a maximum number of dimers on the square lattice (see Fig. \ref{fig1}(c)). The locations of bright spots in specific-heat color maps correlate very well with the variation of transition temperatures with $J_2/J_1$ (see Fig. \ref{fig1}(d)). Thus, the specific heat maps trace out the phase boundaries for the model. We note that for the intermediate all-dimer states, there is no sharp peak in the specific heat, and consequently, a boundary separating the paramagnetic and the all-dimer state is absent. This feature is consistent with the behavior of a quantum paramagnetic ground state. Given that the effective Hamiltonian Eq (\ref{eq:Ham2}) is classical, we can easily compare the energies of the three ground state configurations as a function of $J_2/J_1$.  From this comparison we conclude that the all-dimer state is lower in energy in the range $0.25 < J_2/J_1 < 0.75$. Our simulations also confirm this in the $T \rightarrow 0$ limit. However, thermal disorder favors the presence of magnetically ordered state at low but non-zero temperatures. The order induced by thermal disorder can be easily inferred from the re-entrant feature in the specific heat map (see Fig. \ref{fig1}(d)).

 \begin{figure}[t!]
\includegraphics[width=0.96 \columnwidth,angle=0,clip=true]{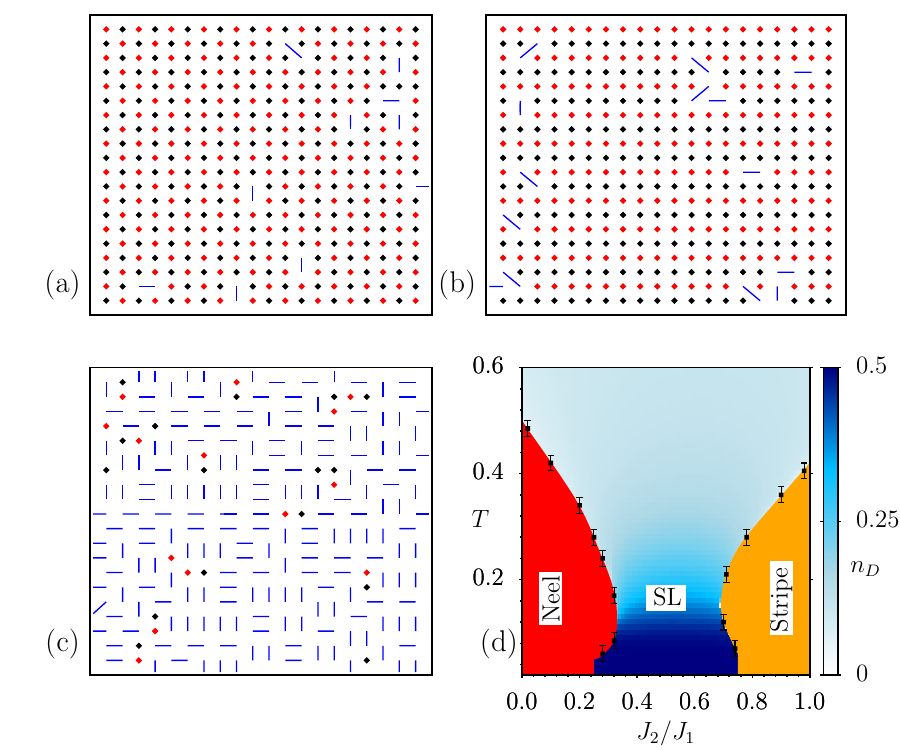}
 \caption{ (Color online) Real-space patterns of spins (red: $s=1/2$, black: $s=-1/2$) and singlets (blue links) of a $20 \times 20$ lattice section taken from a simulation on $ 60 \times  60$ lattice for, (a) $J_2/J_1=0.10$, $T=0.31$, (b) $J_2/J_1=0.90$, $T=0.31$ and (c) $J_2/J_1=0.50$, $T=0.11$. (d) $T - J_2/J_1$ phase diagram inferred from the spin structure factor data. The color bar corresponds to the fraction of dimers, $n_D$, in the SL regime.}
\label{fig2}   
\end{figure}

Representative snapshots from the Monte Carlo simulations are shown for the three qualitatively distinct regimes of the parameter space in Fig. \ref{fig2} (a)-(c). The presence of nn as well as nnn singlets can be noted in the ordered regimes. This suggests that at finite temperatures, the ordered moment can have a reduced value due to the singlet fluctuations. The results can be summarized as a phase diagram (see Fig. \ref{fig2}(d)). Different methods are known to lead to conflicting results in terms of the region of existence, in parameter $J_2/J_1$, of a SL phase. This range is reported to be close to approximately $(0.4, 0.6)$ in most methods \cite{Morita2015, Richter2009, Sushkov2001}. Within our approximation, while the $J_2/J_1$ range for a random-dimer-covering ground state is 0.25 to 0.75 at $T=0$, it narrows down at finite temperatures due to the order-by-thermal-disorder effect. The key argument is that the entropy of independent spin-1/2 variables is larger than that of nn dimer-covering configurations on the square lattice. A similar order-by-quantum-disorder mechanism should reduce the stability range of the spin liquid state at $T=0$ \cite{Schick2020}.  Therefore, we can conclude that our semiclassical approach provides a qualitatively correct picture for the phase diagram of the $J_1 - J_2$ Heisenberg model. Clearly, our approximation is not tailored to 
settle the ongoing debate regarding the range of stability of the SL phase.
Instead, the approach is useful for describing the qualitative features without the need for heavy quantum mechanical calculations. Having tested the performance of our approach for the $J_1-J_2$ model, we now focus on the main issue related to the presence of vacancies.

{\it Effect of spin vacancies:--}
We consider a situation in the $J_1-J_2$ model where a fraction of sites are void of spins. Such a possibility may arise in some strongly insulating magnets upon doping \cite{Glittum2025, Badola2024}. Such models may also be realized in optical lattice settings \cite{Duncan2017}. This leads to a dynamically disordered spin model at the outset that cannot be treated well via methods that rely on translational invariance. The Hamiltonian Eq. {\ref{eq:Ham2}} is easily altered to include the presence of vacancies, and our approach works equally well here. The magnitude of classical spins on the U sites is now redefined as $v_i/2$, where $v_i = 0$ ($1$) represents the presence (absence) of the spin vacancy at the site $i$. Furthermore, an additional constraint, $b_{ij} = 0$ if $v_i v_j = 0$, is implemented. With these conceptually straightforward modifications included in the model and in the Monte Carlo implementation, we present the simulation results for the $J_1-J_2$ model in the presence of vacancies. 

\begin{figure}[t!]
\includegraphics[width=0.96 \columnwidth,angle=0,clip=true]{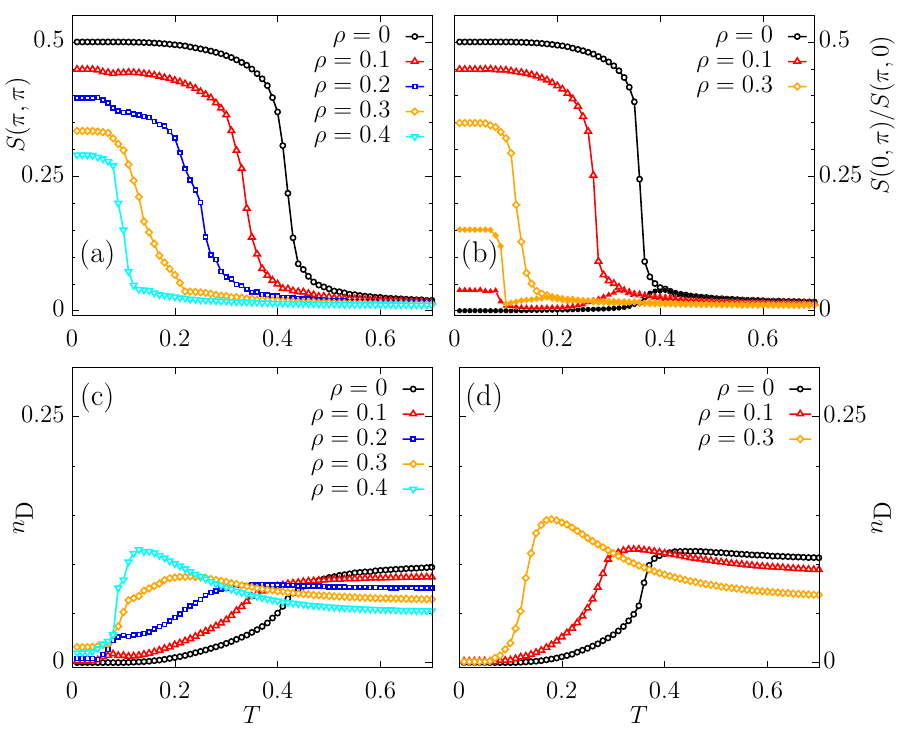}
 \caption{(Color online) Temperature dependence of, (a) $S(\pi,\pi)$ at $J_2/J_1=0.1$, (b) $S(0,\pi)$ (open symbols) and $S(\pi,0)$ (filled symbols) at $J_2/J_1 = 0.9$, the fraction of dimers, $n_\textrm{D}$, at (c) $J_2/J_1=0.1$ and at (d) $J_2/J_1 = 0.9$ for the indicated vacancy concentrations.}
\label{fig3}   
\end{figure}

In Fig. \ref{fig3} (a)-(b), we show the evolution of the order parameters deep inside the N\'eel and stripe ordered phases in the presence of vacancies. The vacancy concentration is denoted by the parameter $\rho$, which is equal to the ratio of the number of missing spins to the total number of sites. Upon increasing $\rho$, the order parameter for the N\'eel phase decreases monotonically (see Fig. \ref{fig3} (a)), in terms of both, the saturation value as well as the onset temperature. Deep inside the stripe phase, we find that either $S(\pi,0)$ or $S(0,\pi)$ increases upon lowering the temperature. The characteristic temperature for this upturn decreases with increasing $\rho$. Interestingly, we observe that the complementary component of the SSF also increases at a lower temperature (see Fig. \ref{fig3} (b)). This leads to ground states with the simultaneous presence of $S(0,\pi)$ and $S(\pi,0)$ with unequal magnitudes.
The corresponding plots for the fraction of dimers display a consistent behavior (see Fig. \ref{fig3} (c)-(d)). \\

\begin{figure}[t!]
\includegraphics[width=0.96 \columnwidth,angle=0,clip=true]{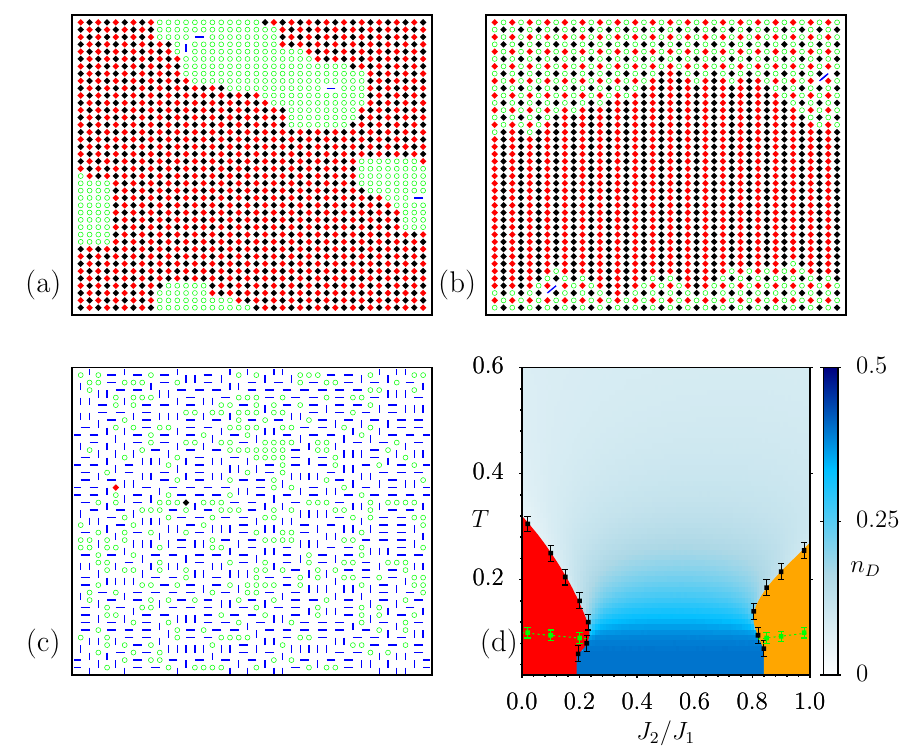}
 \caption{ (Color online) Real-space patterns of spins (red: $s=1/2$, black: $s=-1/2$), singlets (blue links) and vacancies (green circles) on a $40 \times 40$ lattice for $\rho=0.20$ and $T=0.04$ at, (a) $J_2/J_1=0.04$, (b) $J_2/J_1=0.96$, and (c) $J_2/J_1=0.50$. (d) $T - J_2/J_1$ phase diagram inferred from the SSF data at $\rho = 0.20$. The color bar represents the fraction of dimers, $n_\textrm{D}$, and the green symbols mark the vacancy-clustering temperature.}
\label{fig4}   
\end{figure}

In order to understand the behavior of various components of SSF reported in Fig. \ref{fig3}, we plot the typical configurations in the presence of vacancies for $J_2/J_1 = 0.04, 0.50, 0.96$. The low-temperature configurations at $J_2/J_1 = 0.04$ show that the vacancy sites have a tendency to cluster (see Fig. \ref{fig4}(a)). In fact, the on-set temperature for the clustering is lower than that for the $S(\pi,\pi)$. The anomaly near $T \approx 0.1$ (see Fig. \ref{fig3}(c)) in the number of dimers correlates very well with cluster formation. Similarly, for large $J_2/J_1$ we find that the vacancies order in a staggered pattern so that the nnn interactions are partly retained at the cost of nn interactions (see Fig. \ref{fig4}(b)). Note that this vacancy-ordered phase supports the simultaneous presence of both $S(0,\pi)$ and $S(\pi,0)$. While the rise in one of the two components indicates the onset of a stripe magnetic order, the rise in the second component is due to the staggered ordering of the vacancy sites. Deep inside the SL state, the vacancies do not cluster or order even at low temperatures (see Fig. \ref{fig4}(c)). We summarize the discussion above in the form of a phase diagram in the presence of vacancies (see Fig. \ref{fig4}(d)). Comparing with the phase diagram in Fig. \ref{fig2}(d), we find that the intermediate region with no conventional magnetic order is expanded.

\begin{figure}[t!]
\includegraphics[width=0.96 \columnwidth,angle=0,clip=true]{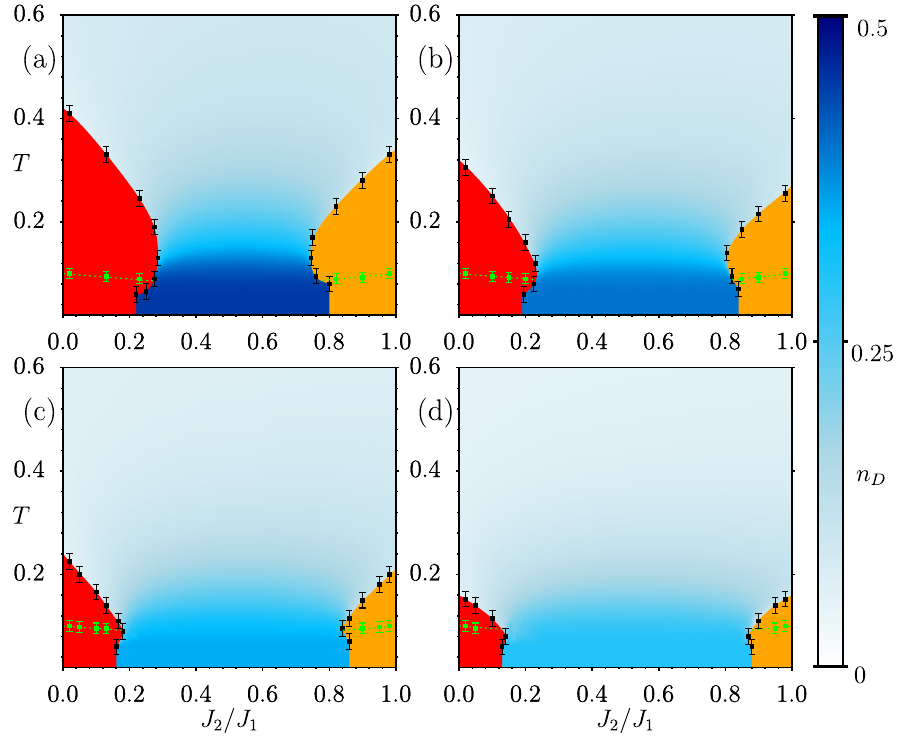}
 \caption{ (Color online) (a)-(d) Evolution of $T - J_2/J_1$ phase diagrams with increasing $\rho$: (a) $\rho=0.1$, (b) $\rho=0.2$, (c) $\rho=0.3$ and (d) $\rho=0.4$.  The color bar represents the number of dimers ($n_\textrm{D}$), and the green symbols denote the vacancy-clustering temperature as inferred from the specific heat.}
\label{fig5}   
\end{figure}

We perform simulations for different values of vacancy concentration, and the results are summarized in the phase diagrams in Fig. \ref{fig5}. We note that the regime with no magnetic order at low temperatures monotonically increases with increasing vacancy concentration. A qualitative understanding for this can be obtained in terms of the higher entropy of the SL state in the presence of vacancies. At low temperatures, clustering of vacancy sites is energetically favored. Hence, there is no additional gain in entropy due to vacancies. On the other hand, in the SL regime, the energy of the system is simply determined by the number of dimers. This allows for a further entropy enhancement in the SL regime. The main challenge in realizing this situation in experiments is in terms of introducing mobile vacancies. Optical lattices could turn out to be a useful platform for a realization of mobile vacancies \cite{yue2022}.

{\it Conclusions and discussion:--}
We have introduced a new semiclassical approach to study the $J_1-J_2$ Heisenberg model in the presence of spin vacancies. The approach is tested for the conventional $J_1-J_2$ model without vacancies, and the results are qualitatively in agreement with those obtained via CMFT and variational MC \cite{Ren2014, Morita2015}.
We find that the SL regime near $J_2/J_1 = 0.5$ is significantly broadened in the presence of mobile vacancies. The enhanced stability of the SL phase is understood in terms of the vacancy contribution to entropy in the SL state. This suggests a new route for finding SL ground states by means of introducing dynamic vacancies in magnetic insulators. One approach could be to tune the frustration index by applying external pressure in doped antiferromagnets. It has been experimentally reported that introducing disorder in magnetic interactions in B-site ordered double perovskites, Sr$_2$Cu(Te$_{1-x}$W$_x$)O$_6$, can induce SL like features \cite{Yoon2021,Mustonen2018a, Watanabe2018, Mustonen2018b}. The key challenge is to identify Mott insulators that resist metallization upon doping \cite{Skolimowsski2019, Yim2024, Glittum2025}. Mott insulators that display phase separation upon doping could also be strong candidates, provided their frustration index is enhanced via internal or external pressure \cite{Yee2015}.
Optical lattices could also provide  a viable alternative platform for a realization of vacancy induced SL phase. It has been shown that mobile spin impurities can be realized on optical lattices \cite{yue2022}.  In addition to the specific results discussed for the  $J_1-J_2$, our approach is generally useful for a semiclassical analysis of other quantum spin models with short-range interactions and disorder. 

{\it Acknowledgments:--}
We acknowledge the use of the HPC facility at IISER Mohali. S.D. acknowledges IISER Mohali for support through the institute fellowship.



\begin{widetext}

\vspace{1cm}


\section*{Supplemental Material}
\setcounter{section}{0}
\setcounter{equation}{0}
\setcounter{figure}{0}
\renewcommand\thesection{\Alph{section}}

\renewcommand{\theequation}{S\arabic{equation}}
\renewcommand{\thefigure}{S\arabic{figure}}

\section{Classical Monte Carlo Simulations}

We perform classical Monte Carlo simulations of the Hamiltonian (described in Eq. (2) in the main text) to investigate the thermodynamic and ground-state properties of the system. Simulations are carried out on square lattices of size $L \times L$ with $L = 40$ and $60$, employing periodic boundary conditions to reduce boundary effects and approximate the thermodynamic limit. The simulations are initialized from a random configuration of classical spins, singlet dimers, and vacancies (when present), generated at a high temperature. The vacancy concentration is fixed for each run and enforced as a global constraint throughout the simulation. There is an additional local constraint that one site can participate in the formation of only one dimer. 

To reach low-temperature equilibrium states, we employ a simulated annealing protocol. The system is gradually cooled down with a sufficient number of discrete temperature steps. At each temperature, $10^5$ Monte Carlo steps (MCS) are used for thermal equilibration, followed by $10^5$ additional steps during which observables are sampled. One MCS consists of $N$ local update attempts, where $N$ is the total number of lattice sites. The update scheme includes multiple local moves:
\begin{itemize}
\item Single-spin flips, which allow magnetic fluctuations.
\item Singlet formation between a spin and an available nearest-neighbor (nn) or next-nearest-neighbor (nnn) partner.
\item Annihilation of existing singlets to restore individual spin degrees of freedom.
\item Spin-vacancy exchange moves, capturing the dynamics of spins in a diluted background.
\item Rotations of singlet bonds to efficiently sample locally resonating singlet configurations.
\end{itemize}

All updates are accepted or rejected according to the standard Metropolis algorithm, based on the energy difference between the current and the proposed configurations. Physical observables such as Energy, Specific heat, singlet densities, and spin structure factors are computed by averaging over Monte Carlo configurations after equilibration. For simulations with finite vacancy density, results are further averaged over several independent runs to improve statistical accuracy.

\section{Real Space Configurations}

\begin{figure}[h]
\includegraphics[width=.8 \columnwidth,angle=0,clip=true]{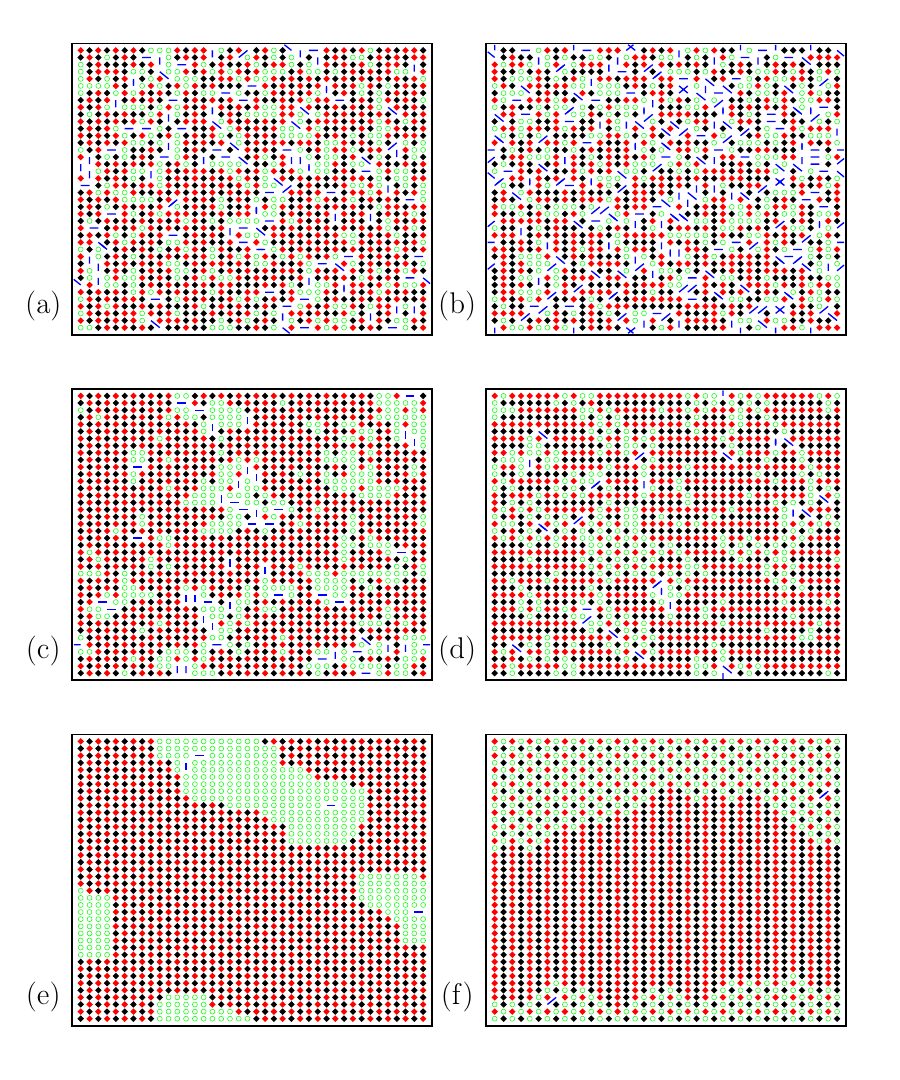}
\caption{Real-space patterns of spins (red: $s=1/2$, black: $s=-1/2$), singlets (blue links) and vacancies (green circles) of a $40 \times 40$ lattice for $\rho = 0.20$ at, $J_2/J_1 = 0.04$, (a) $T = 0.35$, (c) $T = 0.14$, (e) $T = 0.04$ and $J_2/J_1 = 0.96$, (b) $T = 0.35$, (d) $T = 0.14$, (f) $T = 0.04$
}
\label{figS1}
\end{figure}

We present snapshots of the real-space spin and vacancy configurations at a fixed vacancy density of $\rho = 0.20$. Figures \ref{figS1}(a) and \ref{figS1}(b) show typical configurations at a high temperature $T = 0.35$ for $J_2/J_1 = 0.04$ and $J_2/J_1 = 0.96$, respectively, where spins, singlets, and vacancies are distributed randomly. As the system is annealed to an intermediate temperature $T = 0.14$, magnetic order begins to emerge: a N\'eel pattern for $J_2/J_1 = 0.04$ [fig. \ref{figS1}(c)] and a stripe order for $J_2/J_1 = 0.96$ [fig. \ref{figS1}(d)]. However, the vacancies remain disordered at this stage. Upon further cooling to $T = 0.04$, the system develops a distinct vacancy ordering: in the N\'eel phase [fig. \ref{figS1}(e)], vacancies tend to form clusters in the background of antiferromagnetically ordered spins, whereas in the stripe phase [fig. \ref{figS1}(f)], vacancies arrange themselves in a staggered manner. This is reflective of emergent vacancy-vacancy interactions due to magnetic ordering.

\section{Definitions: Spin Structure Factor and dimer fraction}

To characterize magnetic ordering in the system, we have calculated the spin structure factor defined as
\begin{equation}
S(\mathbf{q}) = \frac{1}{N} \sum_{i,j} \langle {S}_i  ~{S}_j \rangle ~ e^{i\mathbf{q} \cdot (\mathbf{r}_i - \mathbf{r}_j)},
\end{equation}
where $N$ is the total number of sites, ${S}_i$ and $\mathbf{r}_i$ denote the spin and position vector at site $i$ whereas $\mathbf{q}$ is the wavevector. In particular, the values at $\mathbf{q} = (\pi, \pi)$ and $\mathbf{q} = (\pi, 0)$ or $(0, \pi)$ are important as these identify the N\'eel and the stripe antiferromagnetic phases, respectively.

In addition, we track the fraction of singlets present in the system at each temperature. This is defined as
\begin{equation}
n_D = \frac{1}{N} \left[ \sum_{\langle  ij \rangle \in C} b_{ij} + \sum_{\langle \langle ij \rangle \rangle \in C} b_{ij} \right],
\end{equation}

\section{Specific Heat}

There is a separation between the onset temperature of the magnetic order and that of the vacancy order. We have calculated the specific heat using the formula $C_V = d\langle E\rangle/dT$. Without any vacancies and $J_2/J_1 \sim 0$, we have only one peak in the specific heat which corresponds to the magnetic phase transition. However, at $J_2/J_1 = 0.30$ as we can see from fig \ref{figS2}(a), there are two peaks. The extra peak, which is at the lower temperature,  corresponds to the reentrant transition. Similar behavior can be observed at $J_2/J_1 = 0.72$, which is shown in Fig. \ref{figS2}(b). For $\rho=0.20$, at $J_2/J_1 = 0.04$, in fig \ref{figS3}(a) we can see two peaks in the specific heat that correspond to magnetic ordering and vacancy clustering. A similar behavior is observed for $J_2/J_1 = 0.96$, which is given in Fig. \ref{figS3}(b). The phase boundary shown in the main text is estimated by the peaks of the specific heat data.


\begin{figure}[h]
\includegraphics[width=.98 \columnwidth,angle=0,clip=true]{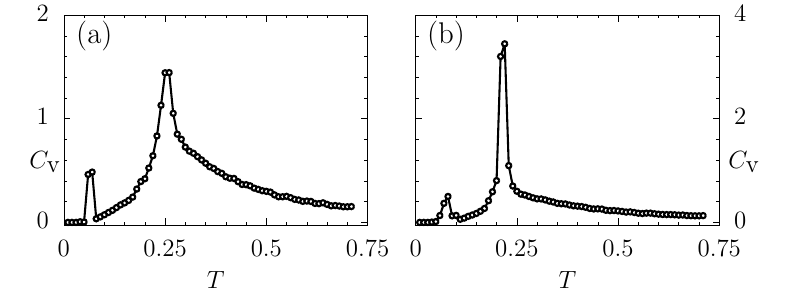}
\caption{Specific heat $(C_{\rm V})$ as a function of temperature for $\rho = 0.0$ (a) $J_2/J_1 = 0.30$ and (b) $J_2/J_1 = 0.72$. 
}
\label{figS2}
\end{figure}

\begin{figure}[h]
\includegraphics[width=.98 \columnwidth,angle=0,clip=true]{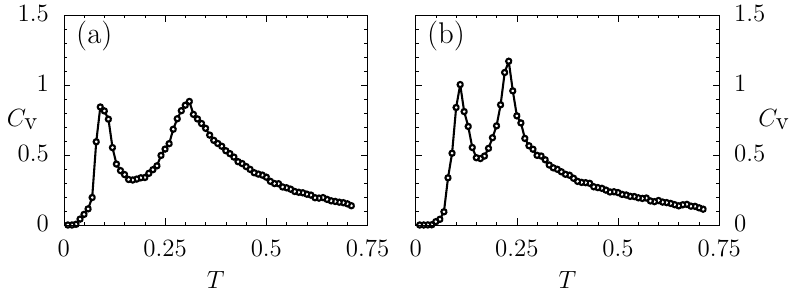}
\caption{Specific heat $(C_{\rm V})$ as a function of temperature for $\rho = 0.20$ (a) $J_2/J_1 = 0.04$ and (b) $J_2/J_1 = 0.96$. 
}
\label{figS3}
\end{figure}

\end{widetext}
\clearpage

\bibliography{main_v1}

\end{document}